\documentclass[%
 reprint,
 amsmath,amssymb,
 aps,pra,twocolumn,
 superscriptaddress,
 longbibliography
]{revtex4-1}

\usepackage{graphicx}
\usepackage{bm}

\renewcommand{\vec}[1]{\mathbf{#1}}

\newcommand{\del}{\partial}
\newcommand{\unit}[1]{\mathrm{\,#1}}

\begin{document}

\title{Collective dipole-dipole interactions in planar nanocavities}
\author{Helge Dobbertin}
\affiliation{Institut f{\"u}r Physik, Universit{\"a}t Rostock, 
Albert-Einstein-Stra{\ss}e 23-24, D-18059 Rostock, Germany}
\author{Robert L{\"o}w}
\affiliation{5. Physikalisches Institut and Center for Integrated Quantum 
Science and Technology, Universit\"at Stuttgart, Pfaffenwaldring 57, 70550 
Stuttgart, Germany}
\author{Stefan Scheel}
\email{stefan.scheel@uni-rostock.de}
\affiliation{Institut f{\"u}r Physik, Universit{\"a}t Rostock, 
Albert-Einstein-Stra{\ss}e 23-24, D-18059 Rostock, Germany}
\date{\today}

\begin{abstract}
The collective response of an atomic ensemble is shaped by its macroscopic 
environment. We demonstrate this effect in the near-resonant transmission of 
light through a thermal rubidium vapor confined in a planar nanocavity. Our 
model reveals density-dependent line shifts and broadenings beyond continuous 
electrodynamics models that oscillate with cavity width and have been observed 
in recent experiments. We predict that the amplitudes of these oscillations can be controlled by coatings that modify the cavity's finesse.
\end{abstract}

\maketitle
When a dense ensemble of quantum emitters is interrogated by near-resonant 
light, strong dipole-dipole interactions cause a collective response of the 
system and alter its spectroscopic properties \cite{dicke54,scully09,guerin16}.
Collective behavior has been studied intensively to build enhanced light-matter 
interfaces such as a perfect mirror out of a single layer of atoms 
\cite{bettles16,shahmoon17} or to control unwanted effects such as collective 
line shifts that can bias optical clocks \cite{chang04,bromley16,campbell17}.
In the past two decades, novel nanophotonic light-matter interfaces were 
developed \cite{chang18} that offer new handles to study and tune collective 
interactions --- especially with respect to two parameters: emitter geometry 
\cite{friedberg73,keaveney12} and modification of the local density of states 
by cavities, waveguides and photonic crystals 
\cite{ruostekoski16,goban15,gonzalez15}.

A prominent example for the geometry-dependence of collective effects is the 
collective Lamb shift (CLS). Friedberg \textit{et al.} \cite{friedberg73} 
predicted that a continuous slab of atoms of thickness $d$ in free space 
exhibits a line shift 
\begin{align}
 \Delta_{\text{CLS}} = \Delta_{\text{LL}}-\frac{3\Delta_{\text{LL}}}{4} 
\left(1- \frac{\sin(2 k d)}{2 k d}\right).\label{eq:CLS}
\end{align}
Here, $\Delta_{\text{LL}}= -\frac{N d_{eg}^2}{3 \epsilon_0 \hbar (2 J_g +1)}$ 
is the Lorentz--Lorenz shift that occurs in bulk media, where $N$ is the atomic 
density, $d_{eg}$ is the transition dipole moment and $(2 J_g +1)$ is the 
number of (degenerate) ground states. Despite its name, the collective Lamb 
shift originates from classical single scattering events (see 
Refs.~\cite{javanainen17,peyrot18} and below). The CLS has been measured in 
x-ray scattering from iron layers in a planar cavity \cite{rohlsberger10},
whereas cold atomic vapor experiments either found a shift compatible with a 
CLS prediction \cite{roof16}, or an altogether negligible shift 
\cite{pellegrino14,jenkins16,corman17}. 

Recently, experiments studied the CLS in wedged nano-cells 
\cite{keaveney12,peyrot18} containing a thermal atomic vapor layer of varying 
thickness $d$ between two planar walls. This confinement intertwines the sample 
geometry with the cavity properties and complicates the evaluation of 
experimental results. Although an earlier study \cite{keaveney12} seemed to 
conform with Eq.~(\ref{eq:CLS}), an improved analysis revealed that it does not 
extend to vapors in cavities and concluded the agreement to be 
fortuitous \cite{peyrot18}. Instead, the latter study found that the 
transmission through the vapor-filled nano-cavity can be described by a 
continuous medium model with an additional broadening and shift that themselves 
depend on density and cavity width.

In this paper, we present a generalized microscopic interaction model that 
consistently accounts for the modified non-collisional atom-atom and atom-wall 
interactions in a planar nanocavity environment. When treating the atomic vapor 
as a continuous medium, we retrieve the well-known transmission profiles of a 
Fabry--Perot etalon. When solving our model for discrete ensembles of thermal 
atoms, the resulting transmission spectra can be fitted to a continuous medium 
model with an additional collective broadening and shift very similar to 
the experimental observations in Ref.~\cite{peyrot18}. We find an oscillatory 
dependence of the collective broadening and shift on the cavity width. The 
amplitudes of these oscillatory patterns can be tuned by coatings that increase 
or decrease the cavity's finesse. This prediction offers a feasible approach 
for experimental tests. Further, such a test also has implications on 
collective interactions in other systems of different shape, dimension, or 
local density of states as the Green's formalism employed here can be adapted 
to any macroscopic environment.

\begin{figure}
 \includegraphics[width=0.99\columnwidth]{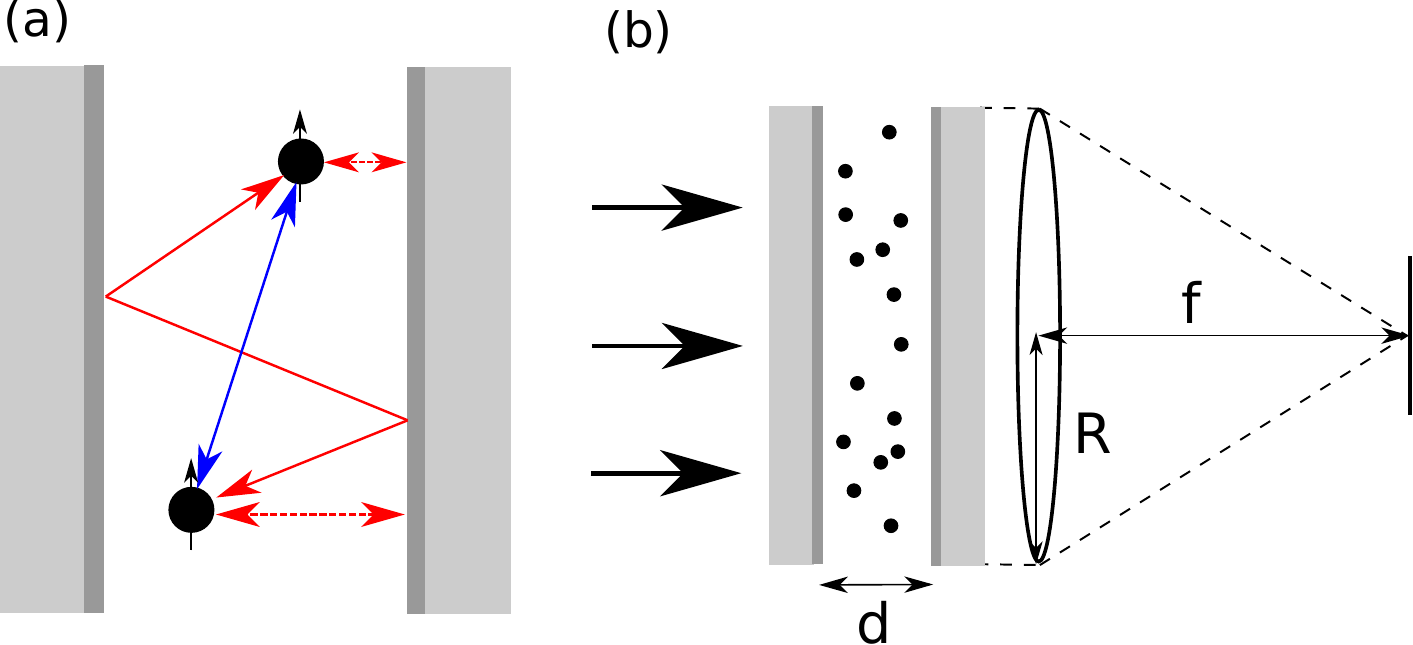}
\caption{(a) Free-space (solid blue line) and cavity-mediated 
(solid red line) atom-atom and atom-wall interactions (dashed lines). 
(b) Sketch of the setup: Incident light is scattered by atoms in a 
cavity of width $d$ and focused onto a detector by a lens of radius $R$. 
\label{fig:Gcav_illustration}}
\end{figure}

The scattering of light in atomic samples is most conveniently described in a 
Green's function formalism, where the classical Green's tensor 
$\bm{G}(\vec{r}_1,\vec{r}_2,\omega)$ is simply the electric field 
at $\vec{r}_1$ radiated by an electric dipole $\vec{d}$ at $\vec{r}_2$, viz. 
$\vec{E}(\vec{r}_1,\omega) = \frac{k^2}{\epsilon_0} 
\bm{G}(\vec{r}_1,\vec{r}_2,\omega) \vec{d}(\vec{r}_2,\omega)$. The Green's 
tensor consists of a free-space and a cavity contribution, 
\begin{align}
\bm{G}(\vec{r}_1,\vec{r}_2,\omega) = 
\bm{G}_{\text{free}}(\vec{r}_1,\vec{r}_2,\omega) + 
\bm{G}_{\text{cav}}(\vec{r}_1,\vec{r}_2,\omega) 
\,. \label{eq:GreenDecomp}
\end{align}
Explicit expressions are provided in the Supplemental Material. If a 
low-intensity field $\vec{E}_\text{inc}(\vec{r})$ impinges onto an atomic ensemble, 
the local steady-state response, $\vec{d}_i = \vec{d}^+_i + \vec{d}^-_i$, follows the classical coupled dipole model (see Supplemental Material) 
\begin{align}
\vec{d}_i^\pm = \bm{\alpha}_i^\pm \vec{E}_\text{inc}^\pm + \bm{\alpha}^\pm_i 
\sum_{j \neq i} \frac{k_0^2}{\epsilon_0} \bm{G}(\vec{r}_i,\vec{r}_j,\omega_0) 
\vec{d}_j^\pm, \label{eq:Coupled_Dipole}
\end{align}
where $\omega_0$ denotes the single-atom resonance frequency in vacuum and $\vec{E}_{\text{inc}}^\pm$ indicates the forward or backward running field components.
The dipole moment of atom $i$ is created by the sum of the incident field in an 
empty cavity, and the field scattered from all other dipoles either directly or 
via the cavity (see Fig.~\ref{fig:Gcav_illustration}a). Without loss of generality, we take a plane wave polarized along $x$ as input field.

For near-resonant light, the atomic polarizability is well described as 
a system with two energy levels $g,e$ with degenerate substates $\mu,\nu$
\begin{align}
\bm{\alpha}_i^\pm = -\frac{1}{\hbar} \sum_{\mu,\nu} \frac{f_{i\substack{g \\ \mu}} 
\vec{d}_{\substack{ge \\ \mu\nu}} \otimes
\vec{d}_{\substack{eg \\ \nu\mu}}}{\left[\delta - 
\Delta_{\substack{ge\\ \mu\nu}}^\pm(\vec{r}_i)
+ \frac{i}{2} \Gamma_{\substack{e \\ \nu}}(\vec{r}_i) \right]}, \label{eq:pola} 
\end{align}
with ground-state population number $f_{i\substack{g \\ \mu}}$, detuning 
$\delta = \omega - \omega_0$, total line shift 
$\Delta_{\substack{ge\\ \mu\nu}}^\pm(\vec{r}) = \pm \omega_{\text{D}} + 
\omega_{\text{CP}\substack{ge\\ \mu\nu}}(\vec{r})$ including the Doppler shift 
$\omega_{\text{D}}=k v$, and the Purcell-enhanced emission rate 
$\Gamma_{\substack{e \\ \nu}}(\vec{r})$ that depends on the dipole's position 
and orientation. The modified emission rate is accompanied by a line shift 
$\omega_{\text{CP}\substack{ge\\ \mu\nu}}(\vec{r})$ deriving from the 
Casimir--Polder interaction between atoms and cavity walls 
\cite{buhmannBook1,buhmannBook2}. We only include the fine structure of the 
atoms. However, Eq.~(\ref{eq:pola}) can be straightforwardly extended to 
include hyperfine states as well \cite{jenkins16}.

From the dipole moment in Eq.~(\ref{eq:Coupled_Dipole}), we compute the 
transmission profile of the sample. Adopting a method established in free space 
\cite{chomaz12}, we consider a lens of radius $R$ (see 
Fig.~\ref{fig:Gcav_illustration}b), operating in the Fraunhofer limit, that 
focuses the forward scattered light onto a detector. The lens integrates over the complex field patterns cast by dipoles and allows a compact expression for the transmission 
coefficient 
\begin{align}
 t &= t_0 + \frac{\sum_j \frac{k^2}{\epsilon_0}\int_{\text{lens}}d 
A_j\,G_{\text{cav,out},xx}(z,\vec{r}_j) d_{x,j} 
}{\int_{\text{lens}}d A\,E_0 e^{i k_1 z}} \nonumber \\
 &= t_0 + \frac{2 i k e^{i d \Delta k}t_{21} }{4 \pi \epsilon_0 R^2} 
\sum_j \frac{e^{-i k z_j}+r_{21} e^{i k z_j}}{1- r_{21}^2 e^{2 
i k d}} \frac{d_{x,j}}{E_0}, \label{eq:transmission}
\end{align}
where $t_0$ is the transmission through an empty cavity, $t_{12}, r_{21}$ are 
the Fresnel coefficients at the interfaces of an empty cavity, $k_1$ is the 
wavenumber in the cavity material, $\Delta k = k-k_1$, $E_0$ is the incident 
field amplitude in free space, and $\bm{G}_{\text{cav,out}}$ is the Green's 
tensor of the light propagating out of the cavity (details of the integration 
can be found in the Supplemental Material).

When treating the atomic vapor as a continuous medium with a spatially 
homogeneous polarizability, our model results in the familiar transmission 
coefficient of a Fabry--Perot etalon. In a high-density vapor, strong line 
shifts and broadenings exceed the single-body Casimir--Polder and Purcell 
effects that do not scale with density. In this regime, it is justified to 
neglect the spatial dependence of the polarizability in Eq.~(\ref{eq:pola}).
Then, the coupled dipole model (\ref{eq:Coupled_Dipole}) results in a field 
governed by new Fresnel coefficients $\tilde{t}_{12},\tilde{r}_{21}$ between 
cavity walls and a gas with refractive index $n_{\text{G}} = \sqrt{1+N 
\alpha_{xx}/\epsilon_0}$. Inserting this field into Eq.~(\ref{eq:transmission}) 
yields the well-known transmission coefficient of a Fabry--Perot etalon
(see Supplemental Material)
\begin{align}
t = \frac{\tilde{t}_{12} \tilde{t}_{21} e^{ i d (k 
n_{\text{G}}-k_1)}}{1-\tilde{r}_{21}^2 e^{2 i n_{\text{G}} k d}} \,.
\label{eq:tMFT}
\end{align}

The collective Lamb shift in Eq.~(\ref{eq:CLS}) derives from the classical Fabry--Perot etalon, Eq.~(\ref{eq:tMFT}), as the limit of a low-density vapor surrounded by free 
space walls \cite{javanainen17,peyrot18}. In this limit Eq.~(\ref{eq:Coupled_Dipole}) can be solved by the single scattering 
(first Born) approximation resulting in a Lorentzian transmission profile that 
features the shift $\Delta_{\text{CLS}}$ (see also 
Refs.~\cite{javanainen17,peyrot18} and Supplemental Material). Deviations from the low-density limit already occur for densities as low as $N= 0.01 k^3$ \cite{javanainen17}. Therefore, Eq.~(\ref{eq:CLS}) cannot be used to study the properties of atomic vapors in the intermediate ($N \sim 1 k^3$) or 
dense ($N\gg 1 k^3$) regimes, even in free space. On the other hand, the 
Fabry--Perot relation (\ref{eq:tMFT}) applies in the dense regime but does not 
account for the discontinuous distribution of atoms and their mutual 
correlations \cite{javanainen17}.  

For a full description, we conducted numerical simulations with discrete 
particles. In each simulation run, we assigned random positions, ground-state 
populations and Doppler shifts sampled from the Maxwell--Boltzmann distribution 
to the otherwise static atoms. Then, we found the self-consistent solution to 
the scattering model (\ref{eq:Coupled_Dipole}) for different detunings $\delta$ 
and computed the transmission spectra according to Eq.~(\ref{eq:transmission}). 
Finally, we computed the mean over many random atomic realizations until the 
transmission profile converged. The transmission profiles can be very well 
fitted to continuum versions of Eqs.~(\ref{eq:Coupled_Dipole}) and 
(\ref{eq:transmission}) that allow for an additional broadening $\Gamma_p$ and 
a line shift $\Delta_p$ that are added to the atomic polarizability 
(\ref{eq:pola}). The fit function also includes non-collisional atom-wall 
interactions and is detailed in the Supplemental Material. Analogously to the 
treatment in Ref.~\cite{peyrot18}, the fitting parameters describe all effects 
that reach beyond the continuous medium model. Repeating the simulations for 
different densities, we found that $\Delta_p$, $\Gamma_p$ scale linearly with 
density. Their slopes $\del_N \Delta_p$, $\del_N \Gamma_p$ describe the 
non-trivial collective effects in the system.

In the simulations, the slab of atoms was truncated to a cylinder of radius 
$R$, with $R=\sqrt{256\pi}/k$ already yielding reasonably well converged fit 
parameters. We conducted the simulations for the D2 line of $^{85}$Rb at room 
temperature with a Doppler width of $\Delta \omega_{\text{FWHM}} \approx 85 
\Gamma_0$. Although the polarizability (\ref{eq:pola}) explicitly involves only 
two energy levels, the Casimir--Polder shift $\omega_{\text{CP}\substack{ge\\ 
\mu\nu}}(\vec{r}) =\Delta \omega_{\mathrm{5P_{3/2}, \nu}}(\vec{r}) -\Delta 
\omega_{\mathrm{5S_{1/2},\mu}}(\vec{r})$ is computed using the full atomic 
multi-level structure. The Casimir--Polder shift scales asymptotically as
$1/R^3$ close to the cavity walls, shifting nearby atoms far away from 
resonance. This results in an inert layer of gas close to the cavity walls 
which gradually changes into a normally responding vapor. Under our conditions 
this normal state is reached about $30\unit{nm}$ away from the surface where the 
Casimir--Polder shift is comparable to the natural linewidth. This spatially 
inhomogeneous response alters the transmission profile through the cavity and 
is included in our fitting function. We performed simulations in a sapphire 
cavity with and without non-collisional atom-wall interactions and found that 
the atom-atom interaction parameters $\del_N \Delta_p$, $\del_N \Gamma_p$ 
remain unchanged within the statistical uncertainty. Hence, non-collisional 
atom-wall interactions do not introduce effects beyond mean-field theory.

\begin{figure}
 \includegraphics[width=0.99\columnwidth]{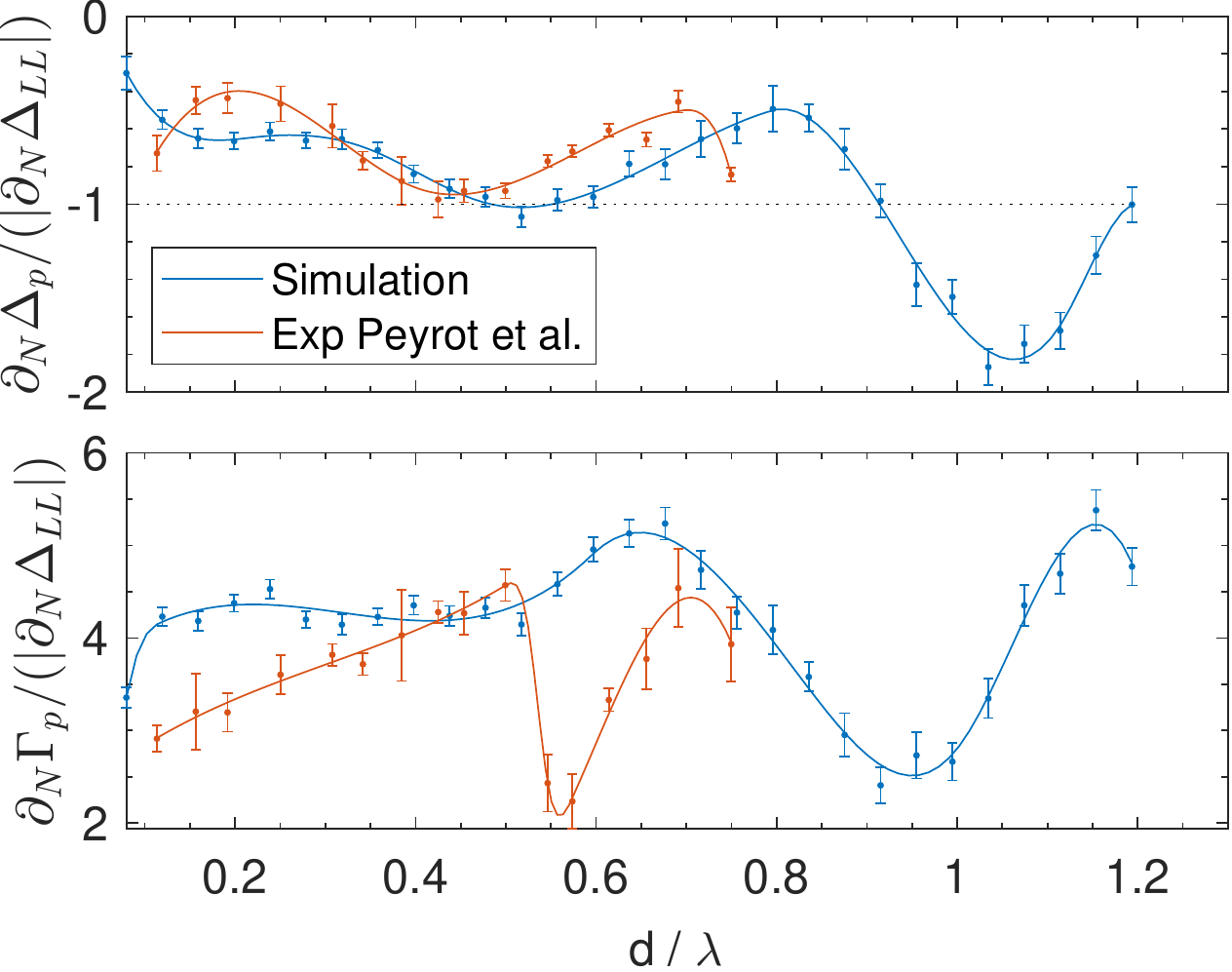}
\caption{Slope of collective shift (top) and broadening (bottom) over width $d$ 
of a vapor-filled sapphire cavity in simulation ($^{85}$Rb, room temperature,
density $\simeq 1 k^3$) and experiment \cite{peyrot18} ($^{39}$K, temperature 
varied up to $600\unit{K}$, density up to $100 k^3$). In textbook 
electrodynamics, a constant Lorentz-Lorenz shift (dotted line) and no 
additional broadening is expected. Simulation error bars correspond to 
1$\sigma$, solid lines are spline interpolations.\label{fig:exp}}
\end{figure}

As a result, we can introduce dimensionless units based on the two-level 
structure of the atoms that describe the generic behavior of any alkali vapor 
by rescaling the cavity size $d$ to the vacuum transition wavelength $\lambda$, 
and the frequency shifts and broadenings to the Lorentz--Lorenz shift 
$\Delta_{\text{LL}}$. Using these dimensionless units, we show in 
Fig.~\ref{fig:exp} the results of the $^{85}$Rb simulation, together with a 
recent experiment \cite{peyrot18} on $^{39}$K vapor where, in both cases, the 
gas is confined in a sapphire cavity. 
The error bars of the simulation show the random error of the fits and 
originate from finite statistics. Whereas the experiment reaches up to atomic 
densities as high as $100 k^3$, our simulations were limited to densities 
around $1 k^3$. This was due to the large numerical effort in solving the 
system of linear equations (\ref{eq:Coupled_Dipole}) that scales as $n^3$ 
with the number of atoms $n$. Further, in the experiment the cell temperature 
is varied to change the density, and reaches higher values ($T=600\unit{K}$) using 
lighter $^{39}$K atoms at the same time.

Despite these differences, simulation and experimental results share key features. The experimentally observed collective shift could be represented by cosine function with a period of $(0.5 \pm 0.02)\lambda$ for a cavity size between $0.1 \lambda$ to $0.75 \lambda$ \cite{peyrot18}. This period approximately matches the series of dips and peaks in the simulation located at $0.52 \lambda$, $0.80 \lambda$ and $1.06 \lambda$ although the shift does not follow simple analytic function like cosine in regions larger than $0.75 \lambda$. The first dip is displaced between simulation and experiment by about $0.1 \lambda$ but features the same minimal value of $\del_N \Delta_p = (-1.0\pm 0.1) \del_N \Delta_{LL}$ (experiment) or $\del_N \Delta_p = (-1.1\pm 0.1) \del_N \Delta_{LL}$ (simulation). The simulated broadening oscillates around the experimental values with quantitative matches in the region between $0.1 \lambda$ and $0.5 \lambda$ cavity width. As a major difference, the experiment shows a pronounced dip in the broadening around $0.55\lambda$ which is not present in the simulation. In sample computations, we found no significant changes to our simulation results when further increasing the temperature. Therefore, the deviations should be considered in the context of four effects that are not covered by the theoretical model.

First, atom-wall collisions change the velocity distribution of the atoms away 
from a Maxwell--Boltzmann shape as investigated in recent experiments 
\cite{todorov17}. Once known, the real velocity distribution should be 
incorporated in the model. Second, the model neglects atom-atom collisions. One 
could discriminate this effect from the collective broadening by investigating 
different atomic lines. Whereas the collisional broadening differs by a factor 
of $\sqrt{2}$ for the D1 and D2 lines \cite{lewis80,weller11}, the 
Lorentz--Lorenz shift, and hence the collective broadening, differs by a 
factor of $2$. Third, a moving atom can absorb a photon in one place and 
radiate at another leading to a nonlocal susceptibility. This introduces a narrowing of the atomic line at low and intermediate densities \cite{peyrot19} including the regime of the simulation  $N=(0.1-1)k^3$. In the dense regime, where the collective effects are probed, the homogeneous broadening is larger than the Doppler width, $\left|\frac{k v}{k d \Gamma}\right| \ll 1$, and the atomic response becomes local \cite{peyrot19}. Therefore, it is consistent to compare the results in Ref.~\cite{peyrot18} to a simulation that does not account for nonlocality. Nonetheless, a full description of experiments conducted at $N \approx 1 k^3$ needs to take into account both collective broadening and shift as well as the nonlocality and should be addressed in a future work. Fourth, we do not account 
for the finite spectrum of the scattered light, and only incorporate the 
Doppler shift to 'zeroth' order \cite{javanainen17} in the polarizability. 
However, modeling moving atoms emitting a finite spectrum of light greatly 
increases the numerical complexity, and no successful implementation of such an 
approach is known to us.

\begin{figure}
\includegraphics[width=0.9\columnwidth]{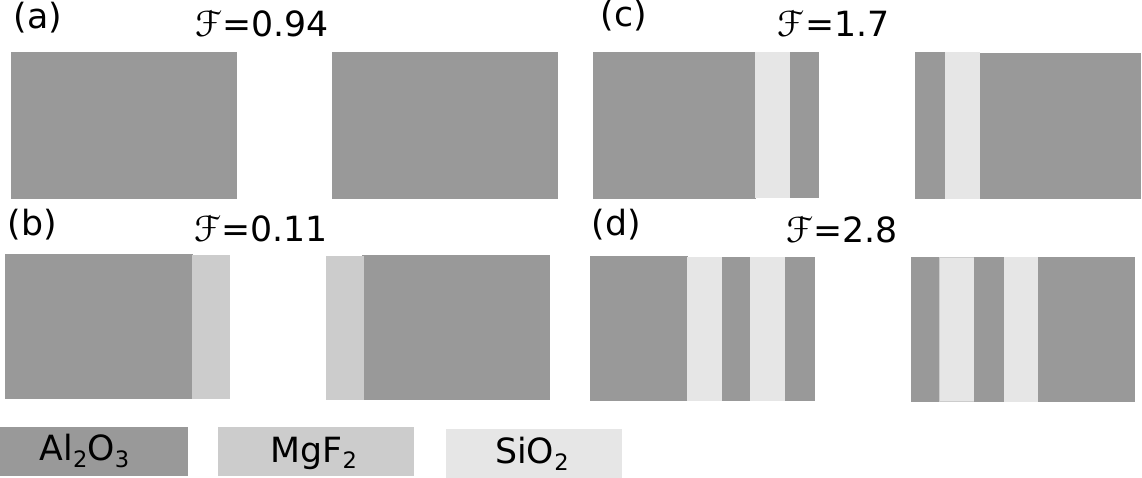}
\caption{Cavity designs of finesse $\mathcal{F}$ with (a) no, (b) 
anti-reflection, or (c),(d) Bragg mirror coating with $\lambda/(4 n)$ 
layers.\label{fig:Coatings}}
\end{figure}

\begin{figure}
\includegraphics[width=0.99\columnwidth]{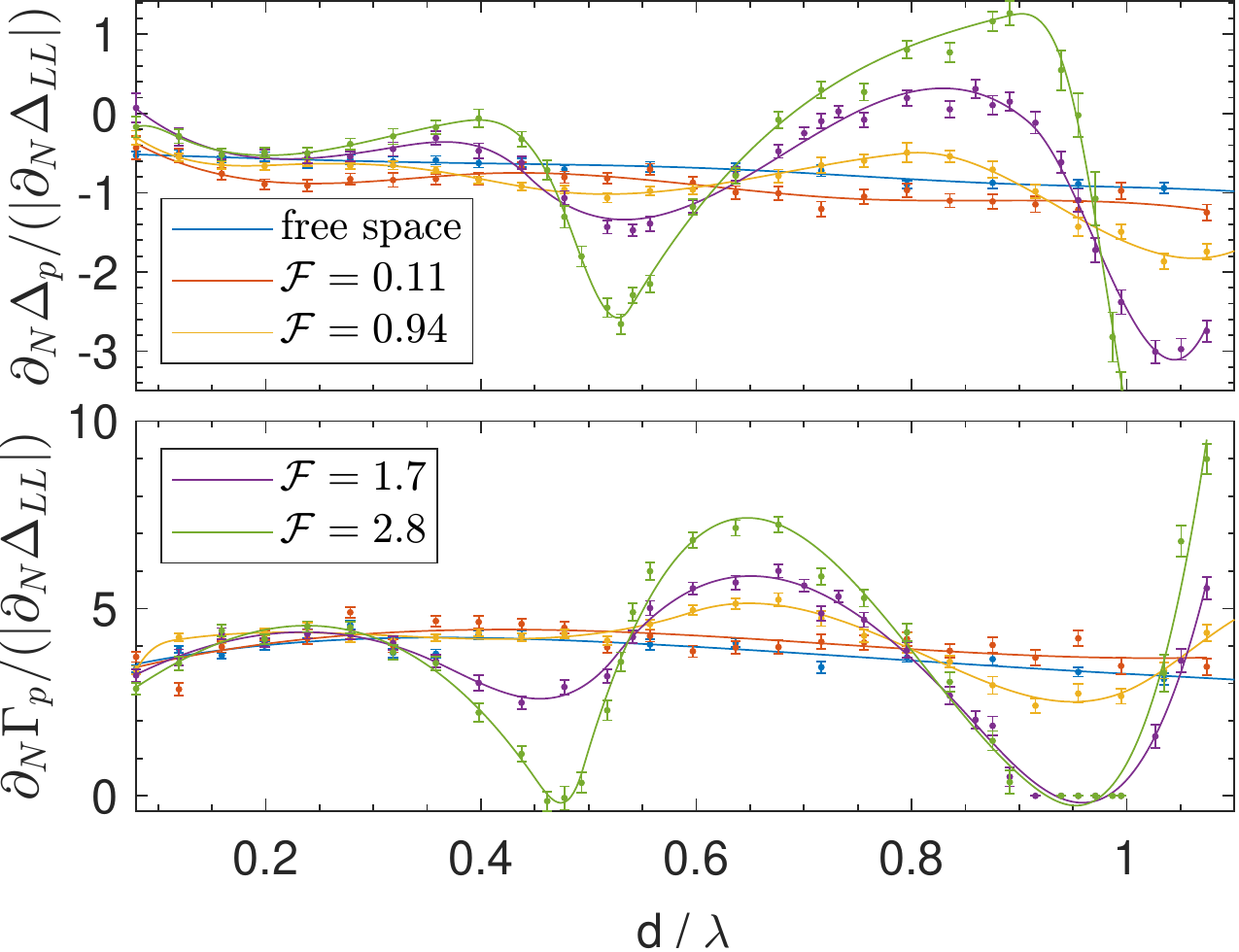}
\caption{Slope of collective shift (top) and broadening (bottom) over width $d$ 
of atomic layer in free space and in the coated cavities depicted in 
Fig.~\ref{fig:Coatings}. Error bars correspond to 1$\sigma$, solid lines are 
spline interpolations.\label{fig:Finesse}}
\end{figure}

We attribute the deviations between simulation and experiment mainly to collisions. Considering the importance of collisions to thermal vapors, it is remarkable that the simulation does not only qualitatively retrieve the observed oscillatory pattern of the collective shift but also quantitatively matches its amplitude. We conclude that modified dipole-dipole interactions play the key role in the behavior of the collective shift and  broadening and propose an experimental design to test this assertion. It 
consists of coated cavities that are equipped with distributed Bragg 
reflectors (DBR) made out of $\lambda/4$ stacks (see 
Fig.~\ref{fig:Coatings}c,d). We used a combination of silica ($n=1.45$) and 
sapphire ($n=1.77$) which provides adequate reflectance.
With increasing cavity finesse we find a prominent enhancement of the 
amplitudes of the oscillatory features in the collective shift 
and broadening as shown in Fig.~\ref{fig:Finesse}. The oscillatory dependence originates from the constructive or destructive interference of photons that mediate correlated atom-atom interactions. Suppression and enhancement of shift and broadening are therefore increased with higher finesse that sharpens the underlying interference effects.
As all cavity designs 
feature the same sapphire surface, the collisional atom-wall interactions and 
the resulting changes to the velocity distribution should be the same. 
Therefore, if future experiments investigate systematic changes in shift and 
broadening for different cavity designs, the results can be compared against 
our prediction even without knowledge about the atom-wall collisions.

Complementary to cavities with large finesse, we considered a cavity with an 
anti-reflection (AR) coating (see Fig.~\ref{fig:Coatings}b). We chose a 
$\lambda/4$ magnesium fluoride ($n=1.38$) layer, which provides decent 
reflection suppression ($R\approx0.1\%$). In practice, one would add a thin 
sapphire layer ($\sim 10\unit{nm}$) on top of the AR coating to protect 
it against the chemically aggressive alkali vapor \cite{ritter18}. The 
collective shift and broadening in the AR-coated cavity shown in 
Fig.~\ref{fig:Finesse} no longer feature pronounced oscillations which can be 
attributed to the fact that the AR coating suppresses photonic interference in the cavity. However, the results for the AR-coated cavity still deviate from 
results obtained in free space. Although the AR coating suppresses reflection 
at normal incidence, the light fields exchanged between atoms also involve 
finite oblique reflection that modify atom-atom interactions.

The presented model is not exclusive to thermal vapor but also applies to 
ultracold atoms. However, one finds that at low temperatures ($T<1\unit{K}$) 
the resulting transmission profiles can no longer be described by continuum 
models, even when introducing more fit parameters such as an effective density. 
As a result, one cannot easily predict or discriminate to which amount a line 
shift is caused by mean-field effects as contained in Eq.~(\ref{eq:tMFT}), or 
by additional collective effects. This may be one reason for the seemingly 
conflicting outcomes of cold vapor experiments mentioned above.

In conclusion, we constructed a microscopic model that accounts for the effects 
of a macroscopic cavity environment on the dipole-dipole interaction between 
atoms. We showed that these modified interactions can explain the oscillatory 
dependence of the collective shift of thermal atoms on the cavity width 
observed in Ref.~\cite{peyrot18}. Furthermore, we predicted that the amplitudes 
of these oscillations can be tuned in coated cavities which can be tested in 
future experiments. The existing literature has explored the atom-atom 
interactions in one- and two-dimensional photonic crystals 
\cite{ruostekoski16,goban15,gonzalez15}, and is now complemented by our 
three-dimensional description. Our model can be used to evaluate the role of
dipole-dipole interactions, Casimir--Polder and Purcell effects in many 
macroscopic environments that have been interfaced with atoms such as 
microresonators \cite{vernooy98}, nanofibers \cite{vetsch10}, hollow-core fibers 
\cite{ghosh06}, superconducting chips \cite{hattermann17}, 
and atomic cladding waveguides \cite{stern13,ritter18}.
The necessary changes have to be incorporated in the respective Green's 
tensor. In a next step, our model should be extended to account for nonlocal 
susceptibilities and saturation effects that are more likely to occur in cavity 
environments. The difficulties to reproduce the bulk Lorentz--Lorenz shift in 
thermal vapors in free space raised questions as to whether the microscopic 
modeling of atom-atom interactions can be considered complete 
\cite{javanainen17}. We hope that future tests to our model can help address 
this fundamental question. 

\begin{acknowledgments}
We thank Charles Adams and Robin Kaiser for fruitful discussions and Tom 
Peyrot for sharing his data and insights on the topic with us.
\end{acknowledgments}

\nocite{Dung02}
\nocite{Buhmann04}
\nocite{Lee16}
\nocite{Paulus00}
\bibliography{cavity_paper}

\end{document}



\title{Supplemental Material: Collective dipole-dipole interactions in planar nanocavities}
\author{Helge Dobbertin}
\affiliation{Institut f{\"u}r Physik, Universit{\"a}t Rostock, Albert-Einstein-Stra{\ss}e 23-24, D-18059 Rostock, Germany}
\author{Robert L{\"o}w}
\affiliation{5. Physikalisches Institut and Center for Integrated Quantum Science and Technology, Universit\"at Stuttgart, Pfaffenwaldring 57, 70550 Stuttgart, Germany}
\author{Stefan Scheel}
\email{stefan.scheel@uni-rostock.de}
\affiliation{Institut f{\"u}r Physik, Universit{\"a}t Rostock, Albert-Einstein-Stra{\ss}e 23-24, D-18059 Rostock, Germany}
\date{\today}

\maketitle
\section{Derivation of the coupled dipole model}
In this section, we derive the coupled dipole model from quantum theory. We start with the atom-field coupling Hamiltonian in dipole approximation \cite{Dung02,Buhmann04}
\begin{align}
\hat{H} = \int d^3 \vec{r} \int_0^{\infty} d \omega\, 
\vecop{f}^{\dagger}(\vec{r},\omega) \cdot \vecop{f}(\vec{r},\omega) \hbar \omega 
+ \sum_A \sum_n E_{A,n} \hat{\sigma}_{A,nn} - \sum_A \int_0^{\infty} d \omega\, 
[\vecop{d}_A \cdot \vecop{E}(\vec{r}_A,\omega) + 
\vecop{E}^{\dagger}(\vec{r}_A,\omega) \cdot \vecop{d}_A],
\end{align}
where $\vecop{d}_A = \sum_{m,n} \vec{d}_{A,mn} \hat{\sigma}_{A,mn}$ and 
$\hat{\sigma}_{A,mn} = \ket{m_A}\bra{n_A}$ are the atomic transition operators. 
The equation of motion for the atomic operators becomes
\begin{align}
\dot{\hat{\sigma}}_{A,mn} = \frac{1}{i \hbar} [\hat{\sigma}_{A,mn},\hat{H}] 
= &-i \omega_{A,nm} \hat{\sigma}_{A,mn} + \frac{i}{\hbar} \int_0^{\infty} d 
\omega\, \Big[ \sum_k (\hat{\sigma}_{A,mk} \vec{d}_{A,nk} - \hat{\sigma}_{A,kn} 
\vec{d}_{A,km}) \cdot \vecop{E}(\vec{r}_A,\omega)
\nonumber \\
&+ \vecop{E}^{\dagger}(\vec{r}_A,\omega) \cdot \sum_k (\hat{\sigma}_{A,mk} 
\vec{d}_{A,nk} - \hat{\sigma}_{A,kn} \vec{d}_{A,km}) \Big], \label{eq:AtomOpEq}
\end{align}
where $\omega_{A,nm} = [E_{A,n}-E_{A,m}]/\hbar$. We expand the electric field 
operator in terms of the Green's function $\bm{G}(\vec{r},\vec{r}',\omega)$ and 
a set of bosonic vector fields $\vecop{f}(\vec{r},\omega)$ describing collective
excitations of the electromagnetic field and the linearly absorbing matter 
\cite{Dung02,Buhmann04}
 \begin{align}
 \vecop{E}(\vec{r},\omega) = i \sqrt{\frac{\hbar}{\pi \epsilon_0}} 
\frac{\omega^2}{c^2} \int d^3 \vec{r}' \sqrt{\Im \epsilon (\vec{r}',\omega)} 
\bm{G}(\vec{r},\vec{r}',\omega) \vecop{f}(\vec{r}',\omega),
\end{align}
and solve its equation of motion in Markov approximation leading to \cite{Dung02,Buhmann04}
\begin{align}
\vecop{E}(\vec{r},\omega,t) = e^{-i \omega t} \vecop{E}(\vec{r},\omega,0) + i 
\mu_0 \sum_A \sum_{m,n} \left(\delta(\omega-\tilde{\omega}_{A,nm}) - 
\frac{i}{\pi} \frac{\mathcal{P}}{\omega-\tilde{\omega}_{A,nm}}\right) \omega^2 
\Im \bm{G}(\vec{r},\vec{r}_A(t),\omega) \vec{d}_{A,mn} \hat{\sigma}_{A,mn}(t).
\end{align}
Here, we introduced the effective frequencies $\tilde{\omega}_{A,nm}$ that 
govern the time evolution of the atomic operators due to the atom-field 
interactions, and that can be determined self-consistently later. 
We insert the field back into Eq.~(\ref{eq:AtomOpEq}) and explicitly denote the 
coherence between degenerate substates $\nu,\mu$ of the energy levels $n,m$ by 
$\hat{\sigma}_{A\substack{nm\\ \nu\mu}}$. Its expectation value becomes
\begin{align}
\braket{\dot{\hat{\sigma}}_{A\substack{ge\\ \mu\nu}}} 
=& i \omega_{A,ge} \braket{\hat{\sigma}_{A\substack{ge\\ \mu\nu}}} + 
\frac{i}{\hbar} \sum_{\substack{k \\ \kappa}} 
\left\{\braket{\hat{\sigma}_{A\substack{gk\\ \mu\kappa}}} 
\vec{d}_{A\substack{ek\\ \nu\kappa}} - \braket{\hat{\sigma}_{A\substack{ke\\ 
\kappa\nu}}} \vec{d}_{A\substack{kg\\ \kappa\mu}}\right\} \cdot
\vec{E}_\text{inc}(\vec{r}_A,t)\nonumber \\
&+\sum_{B} \sum_{\substack{kpq \\ \kappa \delta \epsilon}} 
\Big\{\left(\braket{\hat{\sigma}_{A\substack{gk\\ 
\mu\kappa}}\hat{\sigma}_{B\substack{pq\\ \delta \epsilon}}} 
\vec{d}_{A\substack{ek\\ \nu\kappa}} - \braket{\hat{\sigma}_{A\substack{ke\\ 
\kappa\nu}}\hat{\sigma}_{B\substack{pq\\ \delta \epsilon}}} 
\vec{d}_{A\substack{kg\\ \kappa\mu}}\right) \cdot
\left( -\frac{1}{2}\bmcal{A}_{B,qp}(\vec{r}_A,\vec{r}_{B}) - i 
\bmcal{B}_{B,qp}(\vec{r}_A,\vec{r}_{B}) \right) \vec{d}_{B\substack{pq\\ \delta 
\epsilon}} \nonumber \\
&+\vec{d}_{B\substack{qp\\ \epsilon \delta}} \cdot \left( 
\frac{1}{2}\bmcal{A}_{B,qp}(\vec{r}_{B},\vec{r}_A) - i 
\bmcal{B}_{B,qp}(\vec{r}_{B},\vec{r}_A) \right) 
\left(\braket{\hat{\sigma}_{B\substack{qp\\ \epsilon \delta}} 
\hat{\sigma}_{A\substack{gk\\ \mu\kappa}}} \vec{d}_{A\substack{ek\\ \nu\kappa}} 
- \braket{\hat{\sigma}_{B\substack{qp\\ \epsilon \delta}} 
\hat{\sigma}_{A\substack{ke\\ \kappa\nu}}} \vec{d}_{A\substack{kg\\ 
\kappa\mu}}\right) \Big\},
\end{align}
where we introduced the abbreviations
\begin{align}
\bmcal{A}_{A,nm}(\vec{r},\vec{r}') &= \frac{2 \mu_0}{\hbar} \Theta 
(\tilde{\omega}_{A,nm}) \tilde{\omega}^2_{A,nm} \Im 
\bm{G}(\vec{r},\vec{r}',\tilde{\omega}_{A,nm}), \\
\bmcal{B}_{A,nm}(\vec{r},\vec{r}') &= -\frac{\mu_0}{\hbar \pi} \mathcal{P} 
\int_0^{\infty} d \omega\, \frac{\omega^2 \Im 
\bm{G}(\vec{r},\vec{r}',\omega)}{\omega-\tilde{\omega}_{A,nm}} \nonumber \\
&= -\frac{\mu_0}{\hbar} \Theta (\tilde{\omega}_{A,nm}) \tilde{\omega}^2_{A,nm} 
\Re \bm{G}(\vec{r},\vec{r}',\tilde{\omega}_{A,nm})
-\frac{\mu_0}{\hbar \pi} \int_0^{\infty} d \xi\, 
\frac{\tilde{\omega}_{A,nm}}{\tilde{\omega}^2_{A,nm}+\xi^2} \xi^2 
\bm{G}(\vec{r},\vec{r}',i \xi).
\end{align}
We define the Casimir--Polder line shift and the corresponding Purcell decay rate as
\begin{align}
&\Delta \omega_{A\substack{n \\ \nu\delta}} = \sum_{\substack{k \\ \kappa}} 
\vec{d}_{A\substack{nk \\ \nu \kappa}} \cdot \bmcal{B}_{A,nk}(\vec{r},\vec{r}) 
\vec{d}_{A\substack{kn \\ \kappa\delta}}, \quad
\Gamma_{A\substack{n \\ \nu\delta}} = \sum_{\substack{k \\ \kappa}} 
\vec{d}_{A\substack{nk \\ \nu \kappa}} \cdot \bmcal{A}_{A,nk}(\vec{r},\vec{r}) 
\vec{d}_{A\substack{kn \\ \kappa\delta}}.
\end{align}
Assuming that the modification of the resonance frequency due to interactions 
is small compared to the bare resonance frequency, we can replace 
$\tilde{\omega}_{A,nm} \mapsto \omega_{A,nm}$ in the definition of 
$\bmcal{A}_{A,nm}(\vec{r},\vec{r}')$ and $\bmcal{B}_{A,nm}(\vec{r},\vec{r}')$ 
and compute the line shifts and decay rates perturbatively.

In order to retrieve the classical coupled dipole model, three classical 
assumptions are necessary \cite{Lee16}. First, we assume that we start from and 
remain in an incoherent mixture of ground states 
$\braket{\hat{\sigma}_{A\substack{gg\\ \mu\nu}}} = \delta_{\mu \nu} 
f_{A\substack{g\\ \mu}}$ with occupation numbers $f_{A\substack{g\\ \mu}}$. 
Second, we assume that the ground-state population of atom $A$ is not 
correlated to the coherence of atom $B$, i.e. 
$\braket{\hat{\sigma}_{A\substack{gg\\ \mu\nu}} \hat{\sigma}_{B\substack{ge\\ 
\delta\epsilon}}} = \braket{\hat{\sigma}_{A\substack{gg\\ \mu\nu}}} 
\braket{\hat{\sigma}_{B\substack{ge\\ \delta\epsilon}}}$ when $A\neq B$. Third, 
we assume that the atoms are unsaturated, and hence all excited states are not 
significantly populated. Furthermore, we assume a monochromatic incident field 
with frequency $\omega_L$, i.e. $\vec{E}_\text{inc}(\vec{r},t) = 
\vec{E}_\text{inc}(\vec{r}) e^{-i \omega_L t}$. Then, we transform the equation 
of motion into the rotating frame of the incident field leading to the slowly 
varying coherence amplitude $\braket{\hat{\tilde{\sigma}}_{A\substack{ge\\ 
\mu\nu}}} = e^{i \omega_L t} \braket{\hat{\sigma}_{A\substack{ge\\ \mu\nu}}}$. 
Assuming that the incident light is near-resonant to the $e \leftrightarrow g$ 
transition but far detuned from all other atomic transitions, we discard other 
$\braket{\hat{\sigma}_{B\substack{qp\\ \epsilon \delta}}}$ terms in 
rotating-wave approximation. The effective equation of motion becomes
\begin{align}
\braket{\dot{\hat{\tilde{\sigma}}}_{A\substack{ge\\ \mu\nu}}}
=& i (\omega_L-\omega_{A,eg}) \braket{\hat{\tilde{\sigma}}_{A\substack{ge\\ 
\mu\nu}}} + \frac{i}{\hbar} f_{A\substack{g \\ \mu}} \vec{d}^{*}_{A\substack{ge 
\\ \mu\nu}} \cdot \vec{E}_\text{inc}(\vec{r}_A) \nonumber \\
&+ i\sum_{\delta} (-\Delta \omega_{A\substack{e \\ \nu\delta}} 
+\frac{i}{2}\Gamma_{A\substack{e \\ \nu\delta}}) 
\braket{\hat{\tilde{\sigma}}_{A\substack{ge \\ \mu\delta}}}  
+ i\sum_{\delta} (\Delta \omega_{A\substack{g \\ \delta\mu}} + 
\frac{i}{2}\Gamma_{A\substack{g \\ \delta\mu}}) 
\braket{\hat{\tilde{\sigma}}_{A\substack{ge \\ \delta\nu}}} \nonumber \\
&+ \frac{i \mu_0}{\hbar} \tilde{\omega}^2_{A,eg} f_{A\substack{g \\ \mu}} 
\sum_{A \neq B} \sum_{\delta,\epsilon} \vec{d}_{A\substack{eg \\ \nu\mu}}  \cdot 
\bm{G}(\vec{r}_A,\vec{r}_{B},\tilde{\omega}_{A,eg})  \vec{d}_{B\substack{ge \\ 
\delta\epsilon}} \braket{\hat{\tilde{\sigma}}_{B\substack{ge \\ 
\delta\epsilon}}}.
\end{align}

Different coherences of a single atom $\braket{\hat{\tilde{\sigma}}_{A\substack{ge\\ \mu\nu}}}$, $\braket{\hat{\tilde{\sigma}}_{A\substack{ge\\ \epsilon\delta}}}$ can mutually couple when an atom emits and reabsorbs a photon that in between has changed its polarization after being reflected from a macroscopic body.
In our planar cavity system, the Green's tensor 
$\bm{G}(\vec{r},\vec{r},\omega) = 
\diag(G^{\parallel}(\vec{r},\vec{r},\omega),G^{\parallel}(\vec{r},\vec{r}, 
\omega),G^{\perp}(\vec{r},\vec{r},\omega))$ is diagonal and the z-axis is chosen perpendicular 
to the surface. We take the quantization axis of the atoms along the z-direction and encounter no coupling, i.e. 
$\Delta \omega_{A\substack{k \\ \nu\delta},z}=\Gamma_{A\substack{k \\ \nu\delta},z}=0,$ $\forall \nu \neq \delta$.
When the quantization axis is chosen along the x-direction parallel to the surface, coupling occurs, e.g. $\Delta \omega_{A\substack{e \\ \nu\delta},x} \neq 0$ for $\nu = -3/2, \delta =+1/2$
because $G^{\perp}(\vec{r},\vec{r},\omega)-G^{\parallel}(\vec{r},\vec{r},\omega) \neq 0$. Then, a linear equation system has to be solved. 
Depending on the choice of basis, the individual sublevels may experience different shifts and broadenings. However, the polarizability, which sums over all sublevels, is the same in all bases. 
Introducing the abbreviation $\Gamma_{A\substack{e \\ \nu}} \equiv 
\Gamma_{A\substack{e \\ \nu\nu}}$, the operator equation of motion with a quantization axis in z-direction becomes 
\begin{align}
 \braket{\dot{\hat{\tilde{\sigma}}}_{A\substack{ge\\ \mu\nu}}}
 =& i \left(\omega_L-\omega_{A,eg} -\Delta \omega_{A\substack{e \\ \nu}} + 
\Delta \omega_{A\substack{g \\ \mu}}+ \frac{i}{2} \left[\Gamma_{A\substack{e \\ 
\nu}}+\Gamma_{A\substack{g \\ \mu}}\right] \right) 
\braket{\hat{\tilde{\sigma}}_{A\substack{ge\\ \mu\nu}}} + \frac{i}{\hbar} 
f_{A\substack{g \\ \mu}} \vec{d}^{*}_{A\substack{ge \\ \mu\nu}} \cdot 
\vec{E}_\text{inc}(\vec{r}_A) \nonumber \\
 &+ \frac{i \mu_0}{\hbar} \tilde{\omega}^2_{A,eg} f_{A\substack{g \\ \mu}} 
\sum_{A \neq B} \sum_{\delta,\epsilon} \vec{d}_{A\substack{eg \\ \nu\mu}} \cdot 
\bm{G}(\vec{r}_A,\vec{r}_{B},\tilde{\omega}_{A,eg})  \vec{d}_{B\substack{ge \\ 
\delta\epsilon}} \braket{\hat{\tilde{\sigma}}_{B\substack{ge \\ 
\delta\epsilon}}}.
\end{align}
Casimir--Polder and Purcell effects generally depend on temperature \cite{buhmannBook2}. In our setting the room temperature provides no significant population of the atomic energy levels. Additionally, the separation between atoms and surface never becomes larger than the transition wavelength, and thus remains small compared to the wavelength of thermal photons. Therefore, both the spectroscopic and geometric temperature are low and our zero temperature treatment of dispersion interactions is justified.

The thermal motion of the atoms is accounted for by the hydrodynamic derivative $\frac{d}{dt}\braket{\hat{\tilde{\sigma}}_{A\substack{ge\\ \mu\nu}}} = \frac{\del}{\del t}\braket{\hat{\tilde{\sigma}}_{A\substack{ge\\ \mu\nu}}} + \vec{v} \cdot \nabla \braket{\hat{\tilde{\sigma}}_{A\substack{ge\\ \mu\nu}}}$. The spatial dependence introduced by the
incident field, $\vec{E}_\text{inc}(\vec{r}) = \vec{\tilde{E}}_{\text{inc}}^+ e^{i k z} + \vec{\tilde{E}}_{\text{inc}}^- e^{-i k z}$, can be accounted for by an ansatz $\braket{\hat{\tilde{\sigma}}_{A\substack{ge\\ \mu\nu}}} = \braket{\hat{\tilde{\sigma}}_{A\substack{ge\\ \mu\nu}}^+} e^{i k z} + \braket{\hat{\tilde{\sigma}}_{A\substack{ge\\ \mu\nu}}^-} e^{-i k z}$. The original equation of motion is the sum of two sets of equations for $\braket{\hat{\tilde{\sigma}}_{A\substack{ge\\ \mu\nu}}^\pm}$,
\begin{align}
 \braket{\del_t \hat{\tilde{\sigma}}_{A\substack{ge\\ \mu\nu}}^\pm} + \vec{v} \cdot \nabla \braket{\hat{\tilde{\sigma}}_{A\substack{ge\\ \mu\nu}}^\pm}
 =& i \left(\omega_L-\omega_{A,eg} \mp k v_z -\Delta \omega_{A\substack{e \\ \nu}} + 
\Delta \omega_{A\substack{g \\ \mu}}+ \frac{i}{2} \left[\Gamma_{A\substack{e \\ 
\nu}}+\Gamma_{A\substack{g \\ \mu}}\right] \right) 
\braket{\hat{\tilde{\sigma}}_{A\substack{ge\\ \mu\nu}}^\pm} + \frac{i}{\hbar} 
f_{A\substack{g \\ \mu}} \vec{d}^{*}_{A\substack{ge \\ \mu\nu}} \cdot 
\vec{\tilde{E}}_\text{inc}^\pm \nonumber \\
 &+ \frac{i \mu_0}{\hbar} \tilde{\omega}^2_{A,eg} f_{A\substack{g \\ \mu}} 
\sum_{A \neq B} \sum_{\delta,\epsilon} \vec{d}_{A\substack{eg \\ \nu\mu}} \cdot 
\bm{G}(\vec{r}_A,\vec{r}_{B},\tilde{\omega}_{A,eg})  \vec{d}_{B\substack{ge \\ 
\delta\epsilon}} \braket{\hat{\tilde{\sigma}}_{B\substack{ge \\ \delta\epsilon}}^\pm} e^{\pm i k (z_B-z_A)}.
\end{align}
We assume the local limit, i.e. $\left| \frac{\vec{v}}{d\, \Gamma_{A\substack{e \\ \nu}}} \right| \ll 1$, where $d$ is the thickness of the atomic vapor layer, and neglect the term $\vec{v} \cdot \nabla \braket{\hat{\tilde{\sigma}}_{A\substack{ge\\ \mu\nu}}^\pm}$ which describes that atoms re-emit light at a position different from their original spot of excitation \cite{peyrot19}. In the steady state, $\braket{\del_t \hat{\tilde{\sigma}}_{A\substack{ge\\ \mu\nu}}}=0$, one obtains a linear set of equations for the $\braket{\hat{\tilde{\sigma}}_{A\substack{ge\\ \mu\nu}}^\pm}$. Identifying the 
classical dipole moment with the expectation value $\braket{\tilde{\vecop{d}}_{A,ge}}=\sum_{\mu,\nu} \braket{\hat{\tilde{\sigma}}_{A\substack{ge\\ \mu\nu}}} \vec{d}_{A\substack{ge 
\\ \mu\nu}}$ and introducing $\vec{E}_{\text{inc}}^\pm = \vec{\tilde{E}}_{\text{inc}}^\pm e^{\pm i k z}$, we arrive at Eq.~(3) of the main text.

\section{Explicit form of the Green's tensor for a planar cavity}
The Green's tensor is the unique solution of the Helmholtz equation
\begin{align}
\left[\vec{\nabla}\times \vec{\nabla}\times-k^2\varepsilon(\vec{r}_1,\omega) 
\right] \boldsymbol{G}(\vec{r}_1,\vec{r}_2,\omega) = \bm{\delta}(\vec{r}_1 - 
\vec{r}_2),
\label{eq:Helmholtz}
\end{align}
with the Sommerfeld radiation boundary condition 
$\boldsymbol{G}(\vec{r}_1,\vec{r}_2,\omega) \rightarrow \bm{0}$ when $|\vec{r}| 
\rightarrow \infty$ where $\vec{r} = \vec{r}_1-\vec{r}_2$.
The free-space (bulk) part of the Green's tensor can be expressed as 
\cite{BuhmannBook1}
\begin{align}
 \boldsymbol{G}_{\text{free}}(\vec{r}_1,\vec{r}_2,\omega)
=-\frac{\bm{\delta}(\vec{r})}{3 k^2}+ \frac{e^{i k r}}{4 \pi k^2 r^3} \Big[ 
\left(k^2 r^2+i k r-1\right) \bm{I}
+(3-3 i k r-k^2 r^2) \vec{e}_{\vec{r}} \otimes \vec{e}_{\vec{r}} \Big]. 
\label{eq:GreenFree}
\end{align}
Next, we derive the cavity (scattering) contribution 
$\boldsymbol{G}_{\text{cav}}(\vec{r}_1,\vec{r}_2,\omega)$ in 
Eq.~(2). Due to cylindrical symmetry, we can utilize a 
coordinate system in which the source is located at $\vec{r}_2=(0,0,z_2)$, the 
receiver at $\vec{r}_1=(x,0,z_1)$, with the $z$-axis perpendicular to 
the surfaces. Any other pair of points can be transformed to this set of 
coordinates. By defining the distance $x=\sqrt{(x_1-x_2)^2+(y_1-y_2)^2}$ 
and the angle $\cos \phi = (x_1-x_2)/x$ to the $x$-axis, we can write the 
Green's tensor as
\begin{align}
\bm{G}_{\text{cav}}(\vec{r}_1,\vec{r}_2,\omega) &= 
\bm{G}_{\text{cav}}(x,\phi,z_1,z_2,\omega) = \bm{R}^T(\phi) 
\bm{G}_{\text{cav}}(x,z_1,z_2,\omega) \bm{R}(\phi), \label{eq:Grot}\\
 \bm{R}(\phi) &= \left(\begin{array}{ccc}
\cos \phi & -\sin \phi & 0\\
\sin \phi & \cos \phi & 0\\
0 & 0 & 1
\end{array}\right).
\end{align}

\begin{figure}
\includegraphics[width=0.3\columnwidth]{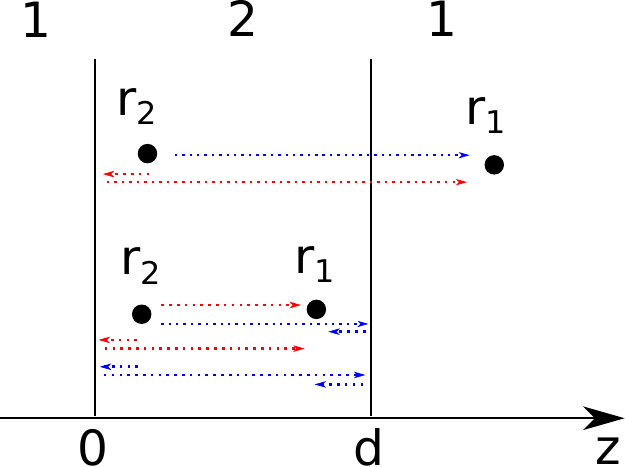}
\caption{Sketch of cavity geometry with photon paths connecting source and receiver points.\label{fig:Gcav_derivation_sketch}}
\end{figure}

Next, $\bm{G}_{\text{cav}}(x,z_1,z_2,\omega)$ is constructed in the basis of 
$s$- and $p$-polarised plane waves \cite{BuhmannBook1}. We 
decompose the wavevector in layer $j$ ($j=1$ outside and $j=2$ inside the 
cavity, see Fig.~\ref{fig:Gcav_derivation_sketch}) into components parallel and 
orthogonal to the surface, $\vec{k}_j = \vec{k}^{\parallel} + 
\vec{k}^{\perp}_j$. Using cylindrical coordinates $\vec{k}^{\parallel} = 
k^{\parallel} (\cos \phi_k, \sin \phi_k, 0)^T$, we write the basis as
\begin{align}
\vec{e}^j_{s\pm} &= \vec{e}_{k^{\parallel}} \times \vec{e}_z = (\sin 
\phi_k,-\cos \phi_k,0)^T, \\
\vec{e}^j_{p\pm} &= \frac{1}{k_j}(k^{\parallel}\vec{e}_z \mp k_j^{\perp} 
\vec{e}_{k^{\parallel}}) = \frac{1}{k_j} (\mp k_j^{\perp} \cos \phi_k,\mp 
k_j^{\perp} \sin \phi_k, k^{\parallel})^T.
\end{align}
Considering the different pathways connecting source and receiver sketched in 
Fig.~\ref{fig:Gcav_derivation_sketch}, we can write down Green's tensors 
analogously to the treatment in Ref.~\cite{BuhmannBook1},
\begin{align}
\bm{G}_{\text{cav}}(x,z_1,z_2,\omega) =& \frac{i}{8 \pi^2} \int_0^{\infty}d 
k^{\parallel}\, \frac{k^{\parallel}}{k^{\perp}_2} \int_0^{2 \pi}d \phi_k\, e^{i 
k^{\parallel} x \cos \phi_k}
 \sum_{\sigma=s,p} \Bigg( \frac{(r_{\sigma}^{21})^2 e^{i k^{\perp}_2 (2 
d+z_1-z_2)}}{D_{\sigma}} \vec{e}^{2}_{\sigma+} \otimes \vec{e}^{2}_{\sigma+} + 
\vec{e}^{2}_{\sigma-} \otimes \vec{e}^{2}_{\sigma+} \nonumber \\
&\times \frac{r_{\sigma}^{21} e^{i k^{\perp}_2 (2 d-z_1-z_2)}}{D_{\sigma}} + 
\frac{r_{\sigma}^{21} e^{i k^{\perp}_2 (z_2+z_1)}}{D_{\sigma}} 
\vec{e}^{2}_{\sigma+} \otimes \vec{e}^{2}_{\sigma-}
+ \frac{(r_{\sigma}^{21})^2 e^{i k^{\perp}_2 (2 d-z_1+z_2)}}{D_{\sigma}} \vec{e}^{2}_{\sigma-} \otimes \vec{e}^{2}_{\sigma-}
\Bigg), \label{eq:GcavInt}
\end{align}
with reflection and transmission coefficients
\begin{align}
r_{s}^{21} &= \frac{k^{\perp}_2-k^{\perp}_1}{k^{\perp}_2-k^{\perp}_1},\quad
r_{p}^{21} = \frac{\epsilon_1 k^{\perp}_2-\epsilon_2 k^{\perp}_1}{\epsilon_1 
k^{\perp}_2-\epsilon_2 k^{\perp}_1}, \label{eq:rCoeffSingle}\\
t_{s}^{21} &= \frac{2 k^{\perp}_2}{k^{\perp}_2 + k^{\perp}_1},\quad
t_{p}^{21} = \frac{k_1}{k_2} \frac{2 \epsilon_2 k^{\perp}_2}{\epsilon_1 
k^{\perp}_2 + \epsilon_2 k^{\perp}_1},\\
D_{\sigma} &= 1-(r_{\sigma}^{21})^2 e^{2 i d k^{\perp}_2}.
\end{align}
For walls consisting of $n-1$ layers, one computes $r_{\sigma}^{21}$ as above 
and infers the effective reflection coefficient of the stack layer by layer 
using the recursion relation
\begin{align*}
 r_{\sigma}^{j+1,j} &= \frac{\tilde{r}_\sigma^{j+1,j}+ r_\sigma^{j,j-1} e^{2 i 
k_j^{\perp} d_j}}{1+ \tilde{r}_\sigma^{j+1,j} r_\sigma^{j,j-1} e^{2 i 
k_j^{\perp} d_j}}.
\end{align*}
Here, $\tilde{r}_\sigma^{j+1,j}$ denotes two-layer coefficients analogously to 
Eq.~(\ref{eq:rCoeffSingle}). The integral over $\phi_k$ in 
Eq.~(\ref{eq:GcavInt}) leads to cylindrical Bessel 
functions $J_n(x)$,
\begin{align}
 J_n(x) = \frac{1}{2 \pi i^n} \int_0^{2 \pi} d \phi\, e^{i x \cos \phi} \cos (n \phi).
\end{align}
The remaining integral over $k^{\parallel}$ can be solved numerically by using 
an appropriate integration contour \cite{Paulus00}.

Furthermore, we encounter integrals of the form $\int d A' 
\bm{G}(\vec{r},\vec{r}')$ that can be computed analytically. Integrating the 
Helmholtz Eq.~(\ref{eq:Helmholtz}) and using the cylindrical symmetry, we obtain
\begin{align}
\int d A' \bm{G}(\vec{r},\vec{r}') &= 
\diag(\tilde{G}^{\parallel}(z,z'),\tilde{G}^{\parallel}(z,z'),0),\\
[\del_z^2+k^2\varepsilon(z,\omega)] \tilde{G}^{\parallel}(z,z') &= - 
\delta(z-z') \label{eq:integratedHelmholtz}.
\end{align}
This is also valid for the Green's tensor $\bm{G}_{\text{cav,out}}(\vec{r},\vec{r}')$ that propagates light from a point inside of the cavity to a point outside of the cavity that is needed in the derivation of Eq.~(5). The one-dimensional Helmholtz equation has the intuitive planar wave solutions (see Fig.~\ref{fig:Gcav_derivation_sketch})
\begin{align}
\int d A'\,\bm{G}_{\text{cav,out}}(z,\vec{r}') &= \diag(1,1,0)\, \frac{i 
t_{21}}{2 k} \frac{e^{i k (d-z') + i k_1 (z-d)}+r_{21} e^{i k (d+z') + i k_1 
(z-d)}}{1- r_{21}^2 e^{2 i k d}},\\
\int d A' [\bm{G}_{\text{free}}(z,\vec{r}')+\bm{G}_{\text{cav}}(z,\vec{r}')] &= 
\diag(1,1,0)\, \frac{i}{2 k} \frac{e^{ i k |z-z'|} + r_{21} e^{ik(z+z')} + 
r_{21} e^{ i k (2 d-z-z')}+r_{21}^2 e^{ i k (2 d -|z-z'|)}}{1-r_{21}^2 e^{2 i k 
d}}.
\end{align}
Note that these results are not limited to the far field.

\section{Continuous medium model}
In this section, we derive the transmission coefficient of the Fabry--Perot 
etalon, Eq.~(6), from our microscopic interaction model. 
After averaging over the Maxwell-Boltzmann distribution, we retain only one polarizability $\bm{\bar{\alpha}}_i = \sqrt{\frac{a}{\pi}} \int_{-\infty}^{\infty} d v\, \bm{\alpha}_i^\pm e^{-a v^2}$ where $a=\frac{m}{2 k_B T}$. As a result the coupled dipole model Eq.~(3) can be rewritten into a single equation for the velocity averaged $\vec{\bar{d}}_i = \vec{\bar{d}}^+_i + \vec{\bar{d}}^-_i$,
\begin{align}
\vec{\bar{d}}_i = \bm{\bar{\alpha}}_i \vec{E}_\text{inc} + \bm{\bar{\alpha}}_i 
\sum_{j \neq i} \frac{k_0^2}{\epsilon_0} \bm{G}(\vec{r}_i,\vec{r}_j,\omega_0) 
\vec{\bar{d}}_j. \label{eq:Coupled_DipoleAveraged}
\end{align}
Averaging over the ground state populations, the polarizability tensor becomes diagonal 
$\bm{\bar{\alpha}}(z)=\diag[\alpha^{\parallel}(z),\alpha^{\parallel}(z),\alpha^{\perp}
(z)]$. The transition from discrete atoms to a continuous gas is accomplished by the 
replacement $\sum_{j\neq i} \rightarrow N \int d V$. We introduce the susceptibility $\chi(z) = N 
\alpha^{\parallel}(z)/\epsilon_0$ and the normalized field 
$\tilde{E}_x(z)=d_x(z)/(\alpha^{\parallel}(z) E_0)$. The coupled dipole model 
Eq.~(\ref{eq:Coupled_DipoleAveraged}) becomes
\begin{widetext}
\begin{align}
\tilde{E}_x(z) =& \tilde{E}_{x,\text{inc}}(z)+ k^2 \int_0^d d z'\, \int d A' 
G_{xx}(z,\vec{r}') \chi(z') \tilde{E}_x(z') \label{eq:continousCDM}\\
\approx& \tilde{E}_{x,\text{inc}}(z)+ k^2 \chi \int_0^d d z'\, \int d A' 
G_{xx}(z,\vec{r}') \tilde{E}_x(z') \nonumber \\
=& t_{12}\frac{e^{i k z}+r_{21} e^{i k (2 d -z)}}{1-r_{21}^2 e^{2 i k d}} + 
\frac{i k \chi}{2} \int_0^d d z'\, \tilde{E}_x(z') \frac{e^{ i k |z-z'|} + 
r_{21} e^{ik(z+z')} + r_{21} e^{ i k (2 d-z-z')}+r_{21}^2 e^{ i k (2 d 
-|z-z'|)}}{1-r_{21}^2 e^{2 i k d}}. \label{eq:ApproxcontinousCDM}
\end{align}
\end{widetext}
In the second step, we neglected the non-collisional atom-wall interactions, 
replacing $\alpha^{\parallel}(z)$ with a constant $\alpha^{\parallel}$. This 
approximation is valid in a regime where density-dependent interactions dominate 
as argued in the main text. In Eq.~(\ref{eq:continousCDM}), we do not 
explicitly account for the singular contribution of the free space Green's 
tensor (\ref{eq:GreenFree}). Its effect can be incorporated by inserting the 
Lorentz--Lorenz shift into $\chi$ and redefining $\tilde{E}_x(z)$ as the locally 
corrected field. The solution of Eq.~(\ref{eq:ApproxcontinousCDM}) then assumes 
the intuitive form
\begin{align}
\tilde{E}_x(z) = \frac{\tilde{t}_{12} (e^{ i k n_{\text{G}} z}+\tilde{r}_{21} 
e^{ i k n_{\text{G}} (2 d-z)})}{1-\tilde{r}_{21}^2 e^{2  i k n_{\text{G}} d}}, 
\label{eq:Econti}
\end{align}
with new Fresnel coefficients $\tilde{t}_{12},\tilde{r}_{21}$ between cavity 
walls and the refractive index $n_{\text{G}} = 
\sqrt{1+N \alpha^{\parallel}/\epsilon_0}$ of the gas. Inserting this result into
the continuous version of Eq.~(5), we obtain the 
transmission profile of the Fabry--Perot etalon
\begin{align}
t =& \frac{t_{12} t_{21} e^{i d (k-k_1)}}{1-r_{21}^2 e^{2 i k d}} +\frac{i k 
\chi}{2} e^{i d (k-k_1)} \int_0^d d z\, \frac{e^{- i k z}+ r_{21} e^{ i k 
z}}{1-r_{21}^2 e^{2  i k d}} t_{21} \tilde{E}_x(z) \label{eq:transmissionconti} 
= \frac{\tilde{t}_{12} \tilde{t}_{21} e^{ i d (k 
n_{\text{G}}-k_1)}}{1-\tilde{r}_{21}^2 e^{2 i n_{\text{G}} k d}}.
\end{align}

\section{Collective Lamb shift \label{sec:LowDensity}}
In this section, we derive the collective Lamb shift of a continuous atomic 
slab in free space. In the low density limit, one can approximate the solution 
of Eq.~(\ref{eq:continousCDM}) by considering only a single scattering event 
(first Born approximation). In free space, the result reads
\begin{align}
\tilde{E}_x \approx \tilde{E}^{(0)}_x + \tilde{E}^{(1)}_x = e^{i k z} + k^2 
\int_0^d d z'\, \int d A' G_{xx}(z,\vec{r}') \chi e^{i k z'}.
\end{align}
The transmission coefficient according to Eq.~(\ref{eq:transmissionconti}) then 
assumes the form
\begin{align}
 t = 1+ \frac{i k \chi d}{2} (1 + \chi \xi), \quad
\xi = \frac{k^2}{d} \int_0^d d z\, \int_0^d d z'\, \int d A' G_{xx}(z,\vec{r}') 
e^{i k (z'-z)}.
\end{align}
We can express the susceptibility $\chi = \frac{3 \Delta_{\text{LL}}}{\delta + 
i/2 \Gamma_0}$ in terms of the Lorenz-Lorentz shift $\Delta_{\text{LL}} = 
-\frac{N d_{eg}^2}{3 \epsilon_0 \hbar (2 J_g +1)}$, where we assumed equally 
populated ground states with $\sum_{\mu,\nu} f_{i\substack{g \\ \mu}} 
\vec{d}_{\substack{ge \\ \mu\nu}} \otimes \vec{d}_{\substack{eg \\ \nu\mu}} = \bm{I} 
\frac{d_{eg}^2}{(2 J_g+1)}$. Employing the Taylor expansion $\frac{1}{1-x} 
\approx 1+x$ for $x \ll 1$, we can approximate the transmission profile by a 
Lorentz curve
\begin{align}
 t &\approx 1 -  \frac{k d \chi}{2 i} \frac{1}{1-\chi \xi} \\
 &= 1- \frac{3 \Delta_{\text{LL}} k d}{2 i (\delta - 3 \Delta_{\text{LL}} \Re \xi) + 6 \Delta_{\text{LL}} \Im \xi - \Gamma_0},
\end{align}
from which we can read off the collective Lamb shift. The resulting expression 
is identical to Eq.~(4.1) in the original derivation of the collective Lamb 
shift by Friedberg \textit{et al.} \cite{Friedberg73}, which reads
\begin{align}
\Delta &= 3 \Delta_{\text{LL}} \Re \xi = \frac{3 \Delta_{\text{LL}} k^2}{d} \Re 
\int_0^d d z\, \int_0^d d z'\, \frac{i}{2 k} e^{i k |z'-z|} e^{i k (z'-z)} \\
&= -\frac{3\Delta_{\text{LL}}}{4} \left(1- \frac{\sin(2 k d)}{2 k d}\right).
\end{align}
Subsequently, Friedberg \textit{et al.} added a term $\Delta_{\text{LL}}$ 
accounting for the local-field correction that stems form the singularity of 
the free-space Green's tensor, Eq.~(\ref{eq:GreenFree}). We obtain the 
collective Lamb shift from an entirely classical theory.

\section{Fitting function \label{sec:fitfun}}
Here, we discuss the fitting function used to extract the parameters 
$\Delta_p$, $\Gamma_p$. Both parameters modify the susceptibility of the vapor 
$\chi(z)$ for which we have to solve a continuous version of the coupled dipole 
model that accounts for non-collisional atom-wall interactions. Abbreviating 
$P_x(z) = \chi(z) \tilde{E}_x(z)$, Eq.~(\ref{eq:continousCDM}) becomes
\begin{align}
\tilde{E}_{x,\text{inc}}(z) = \frac{1}{\chi(z,\Delta_p,\Gamma_p)}P_x(z) - k^2 
\int_0^d d z'\, \int d A' G_{xx}(z,\vec{r}') P_x(z').
\end{align}
The $z'$ integral can be approximated by a Gauss--Laguerre quadrature, and the 
resulting linear set of equations is solved for $P_x(z)$. Eventually, one 
obtains the transmission profile
\begin{align}
 t =& \frac{t_{12} t_{21} e^{i d (k-k_1)}}{1-r_{21}^2 e^{2 i k d}} +\frac{i 
k}{2} e^{i d (k-k_1)} \int_0^d d z\, \frac{e^{- i k z}+ r_{21} e^{ i k 
z}}{1-r_{21}^2 e^{2  i k d}} t_{21} P_x(z).
\end{align}
The susceptibility $\chi(z) = N \alpha^{\parallel}(z)/\epsilon_0$ is computed 
from the ground-state averaged polarizability 
\begin{align}
\bm{\alpha}(z) = \frac{1}{(2 J_g+1)} \sum_{\mu,\nu} \bm{\alpha}_{\substack{ge \\ 
\mu\nu}}(z) = 
\diag[\alpha^{\parallel}(z),\alpha^{\parallel}(z),\alpha^{\perp}(z)].
\end{align}
It contains an average over the Maxwell--Boltzmann velocity distribution 
\begin{align}
\bm{\alpha}_{\substack{ge \\ \mu\nu}}(z) &= -\frac{1}{\hbar} 
\vec{d}_{\substack{ge \\ \mu\nu}} \otimes \vec{d}_{\substack{eg \\ \nu\mu}} 
\sqrt{\frac{a}{\pi}} \int_{-\infty}^{\infty} d v\, \frac{1}{B_{\mu\nu}\mp k v} 
e^{-a v^2} \nonumber \\
&= -\frac{1}{\hbar} \vec{d}_{\substack{ge \\ \mu\nu}} \otimes \vec{d}_{\substack{eg 
\\ \nu\mu}} \frac{\sqrt{\pi a}}{k} \left( D\left(\frac{\sqrt{a} 
B_{\mu\nu}}{k}\right) - i\, e^{-\frac{a B_{\mu\nu}^2}{k^2}} \sign \left(\frac{\Im 
B_{\mu\nu}}{k}\right) \right),
\end{align}
where $D(x) = e^{x^2} \int_0^x d t\, e^{-t^2}$ is the Dawson function, 
$a=\frac{m}{2 k_B T}$ and
\begin{align}
B_{\mu\nu} = \delta-\left[\omega_{\text{CP}\substack{ge\\ 
\mu\nu}}(z)+\Delta_p\right] + \frac{i}{2} \left[\Gamma_{\substack{e \\ 
\nu}}(z)+\Gamma_p \right].
\end{align}

\bibliography{cavity_paper}